# A Point Cloud Enhancement Method for 4D mmWave Radar Imagery

Qingmian Wan, Hongli Peng, *Member, IEEE,* Xing Liao, Kuayue Liu, Junfa Mao, *Fellow,*

*Abstract*—A point cloud enhancement method for 4D mmWave radar imagery is proposed in this paper. Based on the patch antenna and MIMO array theories, the MIMO array with small redundancy and high SNR is designed to provide the probability of high angular resolution and detection rate. The antenna array is deployed using a ladder shape in vertical direction to decrease the redundancy and improve the resolution in horizontal direction with the constrains of physical factors. Considering the complicated environment of the real world with non-uniform distributed clutters, the dynamic detection method is used to solve the weak target sensing problem. The window size of CFAR detector is assumed variant to be determined using optimization method, making it adaptive to different environments especially when weak targets exist. The angular resolution increase using FT-based DOA method and the designed antenna array is described, which provides the basis of accurate detection and dense point cloud. To verify the performance of the proposed method, experiments of simulations and practical measurements are carried out, whose results show that the accuracy and the point cloud density are improved with comparison of the original manufacturer mmWave radar of TI AWR2243.

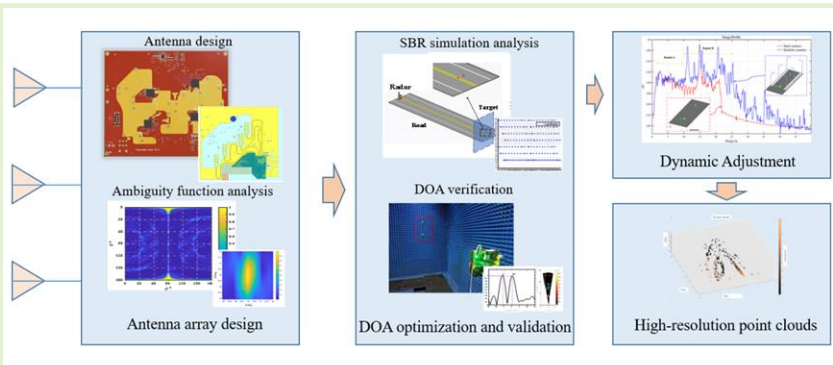

## I. Introduction

WITH the rapid development of the autonomous driving technologies, higher demand of the accurate environment sensing method has been raised [1]. Since the advantage of all-weather working with high resolution, the 4D millimeter wave (mmWave) imaging radar has attracts great interests of researchers [2][3][4][5], which provides the point cloud of targets including the information of range, velocity, azimuth and elevation. Take advantage of the multiple-input multiple-output (MIMO) and time-division multiplexing (TDM) technologies, the 4D mmWave radar synthesizes large number of virtual transceiver channels to acquire sufficient echo data to reconstruct or extract the geometry and velocity information of the targets, such as vehicles, pedestrians, and obstacles. Therefore, the generated point cloud data can be further processed using the methods learned from the camera or Lidar, making the target detection, recognition and tracking easier [6][7][8].

However, to promote the level of autonomous driving and make the vehicle driving safer, the density of 4D mmWave radar point cloud should be further improved. For example, in the multi-target where the radar suffers from the complicated clutters, the 4D mmWave radar faces ambiguity problems such as the separation of targets from environment[9]. Then a coherent phase difference method is proposed in [10] for weak moving target detection in unwanted strong clutter. However, this method cannot deal with the moving object detection when the weak target is not stationary. At the same time, due to the sparsity of the generated point cloud images and the multi-reflection effect in complex environment, it is impossible to distinguish targets using the contexts of the point cloud image[11]. To tackle this problem, a point cloud enhancement method for 4D mmWave radar imagery is proposed in this paper. In the work proposed in this paper, there are four major contributions we have made:

1) Comparison with the commonly used 4D mmWave radar, an optimized antenna design and antenna array deployment method is proposed, which effectively suppresses the sidelobe and shows better SNR (signal-to-noise ratio) performance in object detection with higher angular resolution.
2) Based on the hypothesis of the non-uniform distribution of the environment clutter, an optimized peak spectrum search criterion is proposed in the target detection process of the 4D mmWave radar imagery.
3) The DOA angle resolution, after optimizing the antenna array, creates a more accurate spatial point localization performance, which provides the basis for more precise spatial positioning of the points in the subsequent point cloud.

4) To further analyze the effectiveness of the proposed method, a series of simulations and practical experiments are carried out in detail. Finally, the performances of multi-dimensional radar point cloud imagery such as resolution, point density and spatial accuracy are improved.

The rest of this paper is organized as follows. Section II describes the basic principles of the 4D mmWave radar imagery in the detail view of the signal processing. In section III, a point cloud enhancement method for 4D mmWave radar imagery is studied, including the antenna design, peak spectrum optimization and DOA estimation. Section IV shows the experiment results including the simulation and the self-developed radar prototype which verify the efficiency of the proposed method. Section V gives the conclusion of this paper.

## II. Basic Principles Of The 4D MmWave Radar Imagery

The mmWave radar point cloud imagery diagram is shown in Figure 1. Firstly, a 2D fast Fourier transform (FFT) is performed on the recorded echo data to get the range-Doppler map (RD map) of the detected target. Secondly, the generated RD maps from different transceiver channels are summed together to improve the SNR level and the CFAR detection is taken to obtain the target position in the RD map. Thirdly, the azimuth and elevation angles are estimated using DOA method typically the 2D FFT method. Finally, the 4D point cloud of the target is generated by converting the detected range, velocity, azimuth angle and elevation angle from spherical coordinate into Cartesian coordinate.

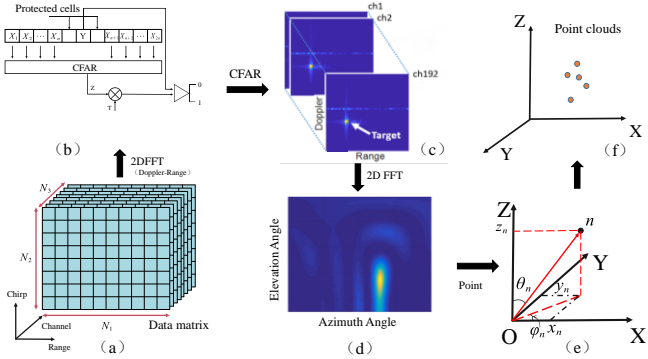

Fig. 1. The mmWave radar point cloud imagery diagram

### A. RD map generation

Generally, the 4D mmWave Radar emits frequency modulated continuous wave (FMCW) signal with $M$ different transmitters using TDM, which can be written as[12]:

$$s_m(t) = \exp(j2\pi f_c t + j\pi k_r t^2) \quad (1)$$

where $f_c$ is the carrier frequency, $k_r$ is the chip rate of the transmitted signal.

Derived from the scattering center hypothesis, the detected target can be modeled as $Q$ scatterers. Then the echo signal scattered from the $q$-th scatterer between the $m$-th transmitter and the $n$-th receiver at time $t_a$ can be written as:

$$\begin{aligned} s_{mn,q}(t,t_a) = &\ \sigma_q \\ &\cdot \exp\left(j2\pi f_c\left(t - \frac{R_{mn,q}(t_a)}{c}\right) \right. \\ &\left. + j\pi k_r\left(t - \frac{R_{mn,q}(t_a)}{c}\right)^2\right) \end{aligned} \quad (2)$$

where $\sigma_q$ is the scattering coefficient of the $q$-th scatterer, $R_{mn,q}(t_a)$ is the sum of the distances from the $q$-th scatterer to the $m$-th transmitter and the $n$-th receiver respectively, and $c$ is the light velocity. Then the total echo signal from the scenario between the $m$-th transmitter and the $n$-th receiver can be written as:

$$\begin{aligned} s_{mn}(t,t_a) = &\sum_{q=1}^{Q} \sigma_q \\ &\cdot \exp\left(j2\pi f_c\left(t - \frac{R_{mn,q}(t_a)}{c}\right) \right. \\ &\left. + j\pi k_r\left(t - \frac{R_{mn,q}(t_a)}{c}\right)^2\right) \end{aligned} \quad (3)$$

The echo signal is detected and sent to the mixer where the difference-frequency of the transmitted signal is generated, after arrangement which can be written as:

$$\begin{aligned} \bar{s}_{mn}(t,t_a) = &\sum_{q=1}^{Q} \sigma_q \cdot \exp\left(j\frac{2\pi R_{mn,q}(t_a)}{\lambda}\right) \\ &\cdot \exp\left(-j2\pi k_r\left(\frac{R_{mn,q}(t_a)}{c}\right)t\right) \\ &\cdot \exp\left(j\pi k_r\left(\frac{R_{mn,q}(t_a)}{c}\right)^2\right) \end{aligned} \quad (4)$$

The last phase of (4) is called residual video phase which can be neglected in non-focused imagery such as the RD map generation. Then the Fourier transform is performed in range direction to get the target profile which can be written as

$$\begin{aligned} \bar{s}_{mn}(f,t_a) = &\ T_p \cdot \sum_{q=1}^{Q} \sigma_q \cdot \exp\left(j\frac{2\pi R_{mn,q}(t_a)}{\lambda}\right) \\ &\cdot \operatorname{sinc}\left(T_p\left(f - 2k_r\frac{R_{mn,q}(t_a)}{c}\right)\right) \end{aligned} \quad (5)$$

where $T_p$ the chirp signal pulse duration, $f$ is the modulated frequency of the baseband signal and $\lambda$ is the wave length.

Subsequently, the second Fourier transform is performed in Doppler direction to get the target velocity. Without loss of the generality, $v_{r,q}$ is assumed the radial velocity between the $q$-th scatterer and the radar. Then $R_{mn,q}(t_a)$ can be represented as $R_{mn,q}(t_a) = R_{mn,q}(t_{q_0}) - v_{r,q} t_a$ and the Fourier transformation of (5) can be written as:

$$\bar{s}_{mn}(f, f_d) = T_a \cdot T_p$$
$$\cdot \sum_{q=1}^{Q} \sigma_q$$
$$\cdot \exp\left(-j2\pi \frac{R_{mn,q}(t_{q_0})}{\lambda}\right) \quad (6)$$
$$\cdot \mathrm{sinc}[T_a(f_d - f_q)]$$
$$\cdot \mathrm{sinc}\left[T_p\left(f - 2k_r \frac{R_{mn,q}(t_a)}{c}\right)\right]$$

where $f_q$ is the Doppler frequency caused by the target movement and $f_q = 2v_r/\lambda$. $T_a$ is the coherent processing time during the target movement.

### B. CFAR detection

CFAR is the technology to determine the presence of the target signal while maintaining a constant false alarm probability by distinguishing the target signal and the noise. It compares the signal to a threshold, which is a function of both the probabilities of the detection and the false alarm, for judging whether the target exists. The widely used cell averaging CFAR (CA-CFAR) detector extracts noise samples from both leading and lagging cells, which are called training cells, around the cell under test (CUT)[13], as shown in Figure.2.

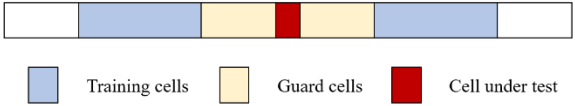

Fig. 2. Schematic of the CA-CFAR detector.

The guard cells are placed adjacent to the CUT, which is used to avoid signal components from leaking into the training cells. The average value of the training cells is computed as the noise estimate. The threshold factor is given as

$$a = N_{tc} \cdot \left(P_{fa}^{-1/N_{tc}} - 1\right) \quad (7)$$

where $N_{tc}$ is the number of the training cells and $P_{fa}$ is the desired false alarm rate.

### C. DOA estimation

After the CFAR operation, $Q$ scatterers are detected with information of range, velocity and their values in different RD maps from $M$ by $N$ channels. The MIMO array provides number of channels to obtain information matrix in azimuth and elevation dimension of the target, which can be used to estimate the direction of arrival (DOA) angles[14]. Simplify the DOA problem into one dimension in azimuth, the received signals of the $q$-th scatterer, ignoring the constants, can be written as

$$a(l|q) = \{\bar{s}_{mn}(f, f_d|q)\}$$
$$= \left\{\exp\left(-j2\pi \frac{R_{mn,q}}{\lambda}\right)\right\}, m \quad (8)$$
$$= 1,2, \ldots, M; n = 1,2, \ldots, N.$$

with the stationary suppose, which means $t_{q_0}$ is fixed.

From the phase center approximation (PCA) algorithm, $R_{mn,q}$ can be modified using the approximated phase center as $R_{mn,q} = 2R_{c_{mn},q}$. The approximated virtual center geometry of the MIMO antennas is always elaborately designed with uniform distribution in one direction.

When the virtual array is uniform placed, the space between each adjacent antenna is represented as $d$. Assuming the $q$-th scatterer is located at the direction of angle $\theta_q$ based on the far-field hypothesis, the received signals of the entire virtual array can be written as

$$a(\theta_q) = \exp\left(-j2\pi \frac{(l-1)d \sin \theta_q}{\lambda}\right), \quad (9)$$
$$l = 1,2, \ldots, MN.$$

Then the estimated target angle can be derived from the object function of DOA, using the steering vector of the virtual array, can be written as

$$\tilde{\alpha}_q = \arg\max_{\alpha_q} \left(a(\alpha_q)^H \cdot a(\theta_q)\right) \quad (10)$$

where

$$a(\alpha_q) = \exp\left(-j2\pi \frac{(l-1)d \sin \alpha_q}{\lambda}\right), l \quad (11)$$
$$= 1,2, \ldots, MN.$$

Therefore, (10) can be rewritten as

$$\tilde{\alpha}_q = \arg\max_{\alpha_q} \left(\sum_{l=1}^{L} \exp\left(j2\pi \frac{(l-1)d(\sin \alpha_q - \sin \theta_q)}{\lambda}\right)\right) \quad (12)$$

with $L = MN$.

## III. THE POINT CLOUD ENHANCEMENT METHOD

In order to enhance the performance of the 4D mmWave radar point cloud imagery, including the SNR, the angle estimation accuracy and the density of the points, the antenna and antenna array are elaborately designed which lead to more accurate azimuth angle measurement, and the detection method is also optimized in un-uniform distributed clutter environment.

With the incensement of the number of the MIMO transceiver antennas, the virtual antenna elements are increased significantly. The desired function of the 4D mmWave radar is then upgraded and promised to better reconstruction of the target details including distributions and contours, while the conventional radar still follows the algorithm of target detection. The 4D mmWave radar needs the cooperation of new detection algorithms, which can better restore the contour information, spatial location and other kinds of detailed features of objects, thus laying a good foundation for realizing object recognition. The detailed work we have made effort in these problems will be described in this section.

### A. Antenna and antenna array design

Based on the principle of 4D mmWave radar imagery, the antenna and antenna array design are important for the whole imagery system. Therefore, the antenna and antenna array design are firstly carried out, theoretically and practically.

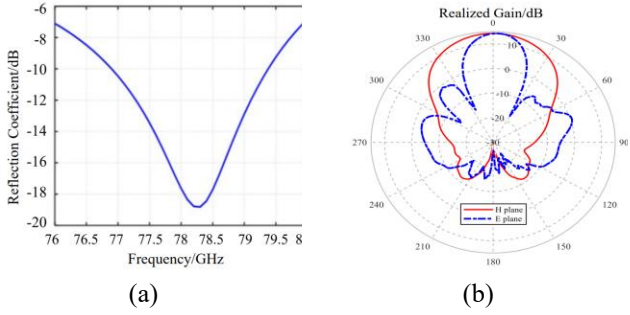

Fig. 3. The performance of the antenna design, (a) reflection coefficient of simulation, and (b) realized gain.

In the design of the antenna array in pitch direction, the pattern of the virtual array can be regarded as the product of the transmit array and the receive array. A wideband micro-strip patch antenna is simulated with optimization and realized, whose performance is shown in Figure 3. Therefore, the beamwidth of the main-lobe is designed narrow but not beyond the system light-of-view (LoV), and the noise from the gate-lobe can be reduced. As shown in Figure 3(b), the pitch resolution of the radar system can be further improved by increasing the number of virtual channels.

To further improve the radar horizontal resolution, the technology of minimum redundancy array [15] is adopted in the design of the antenna array in azimuth direction. The redundancy is used to characterize the aperture utilization of the array, which is positive to the repetition rate of the cross-correlation function (CCF) of the received signals within the target array. The optimization design of the array redundancy can reduce the number of the antennas used. Then the minimum redundancy array is the target of the optimization design.

Suppose the locations of the virtual transceivers of the MIMO array are $d_1, d_2, \ldots, d_L$ and $L = MN$. The possible DOAs of the target can be written as:
$$\boldsymbol{\varphi} = \{\varphi_q | 0 \leq \varphi_q \leq \pi, i = 1,2, \ldots, Q\} \quad (13)$$
The recorded echo signals of the array can be written as:
$$\boldsymbol{x} = \boldsymbol{A}\boldsymbol{s} + \boldsymbol{n} \quad (14)$$
where $\boldsymbol{n}$ is the noise vector, $\boldsymbol{s}$ is the vector of the echo signals from different DOAs, and $\boldsymbol{A}$ is the steering matrix of the array with
$$\boldsymbol{A} = \begin{bmatrix} a_{11} & \cdots & a_{1Q} \\ \vdots & \ddots & \vdots \\ a_{L1} & \cdots & a_{LQ} \end{bmatrix} \quad (15)$$
where $a_{lq} = \exp\left(-j\frac{2\pi d_l}{\lambda}\sin\varphi_q\right)$. Then the covariance matrix of the received signals can be written as:
$$\boldsymbol{R} = \mathrm{E}[\boldsymbol{x}\boldsymbol{x}^{\mathrm{H}}] = \boldsymbol{A}\boldsymbol{s}\boldsymbol{s}^{\mathrm{H}}\boldsymbol{A}^{\mathrm{H}} + \mathrm{E}[\boldsymbol{n}\boldsymbol{n}^{\mathrm{H}}] \quad (16)$$
Suppose the echo signals are incoherent from different DOAs, which can be described as:
$$\boldsymbol{s}\boldsymbol{s}^{\mathrm{H}} = \mathrm{diag}(P_1, P_2, \ldots, P_Q) \quad (17)$$
where $P_q$ is the power of the echo signal. The noise $\boldsymbol{n}$ is assumed in Gaussian distribution with mean value of zero. Then the covariance matrix of the signals can be arranged and written as:
$$\boldsymbol{R} = \{R_{ij}\}, i = 1,2, \ldots, L; j = 1,2, \ldots, L. \quad (18)$$
with $R_{ij} = \sum_{k=1}^{Q} P_k \cdot \exp\left(-j\frac{2\pi(d_i - d_j)}{\lambda}\sin\varphi_k\right)$.

It can be seen from (18) that the covariance of signals depends on the phase difference of the array elements, which is caused by the antenna positions of the array design. Two array elements with the fixed distance have the same covariance, which there is redundancy between the two elements. Consider the array with $L$ uniform distributed elements, the number of the distance differences of each two elements is $L - 1$, which means there are $L - 1$ unique values in the covariance matrix $\boldsymbol{R}$ and $L^2 - L + 1$ pairs of the array elements are redundant. Then the definition of the array redundancy is introduced as:
$$\mathcal{R} = \frac{C_L^2}{\mathcal{L}_{\max}} \quad (19)$$
where $C_L^2$ is the combination number of $L$ and $\mathcal{L}_{\max}$ is the number of the unique values in the covariance matrix. Obviously, the smaller $\mathcal{R}$ means higher usability of the array elements.

However, limited by the factors such as array physical size and feeder layout, it is hard to achieve the minimum redundancy of the MIMO array in practice system design. Since reducing the redundancy has a significant improvement of the MIMO array efficient, optimization of the array design to minimize the redundancy is performed.

In our design of the MIMO antenna array, the constraints from the practice usage of the autonomous driving are determined as follows:
1) The FoV in horizontal is not less than ±60 degrees and in vertical not less than ±20 degrees.
2) The angular resolution in horizontal is not greater than 2 degrees and in vertical not greater than 6 degrees.
3) The physical size of the array is not larger than $35\lambda \times 35\lambda$.
4) The number of transmitter antennas is 12 and the receiver antennas 16.

Considering the impact of the above indicators on array design, a 2D MIMO array layout with ladder shape is proposed in this paper. The ladder shape is a quasi-plane array scheme proposed to form the approximate uniform array. Due to the high integration of mmWave radar, the antenna array plane is usually deployed on the same multi-layer PCB with several other components. It means that the positions of the antenna array elements inevitably need to compromise with other key circuit structures, and the complete 2D MIMO plane array is difficult to be achieved. Therefore, the ladder shape array is one of the suitable deployment schemes for miniaturized MIMO mmWave radar, bring higher degrees of freedom to the deployment of array elements. The final MIMO array is designed with ladder shape and shown in Figure 4.

The receiver elements are distributed in a ladder shape, while the transmitter elements are arranged in a minimum redundant line array. Therefore, the synthesized virtual array is approximate a uniform array, which can effectively suppress the noises from sidelobes of the echo signal. The sparse array is introduced in the vertical direction which achieves approximate the performance of the complete virtual array using the 7 rows array with smaller redundancy.

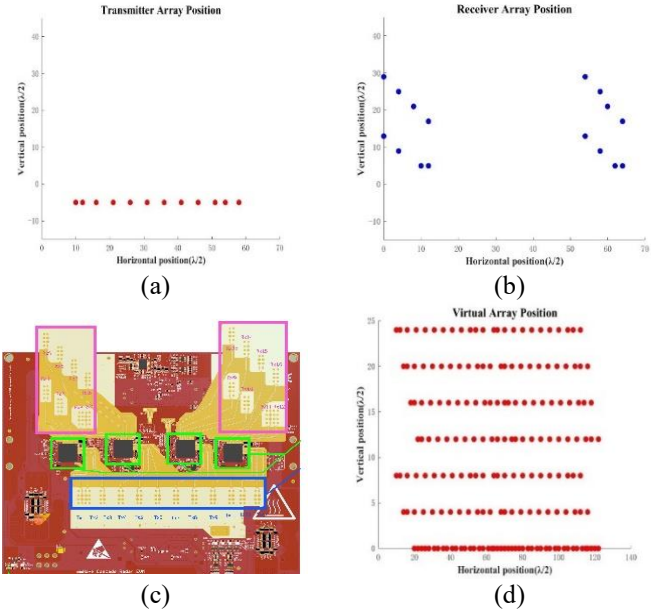

(a) (b)

(c) (d)

Fig. 4. The designed ladder shape array with, (a) the transmitter antenna array, (b) the receiver antenna array, (c) the physical MIMO array and (d) the final virtual array.

The ambiguity function[16] is introduced to analyze the resolution of the 4D mmWave radar system with designed antenna array, which is computed using simulation method and shown in Figure 5.

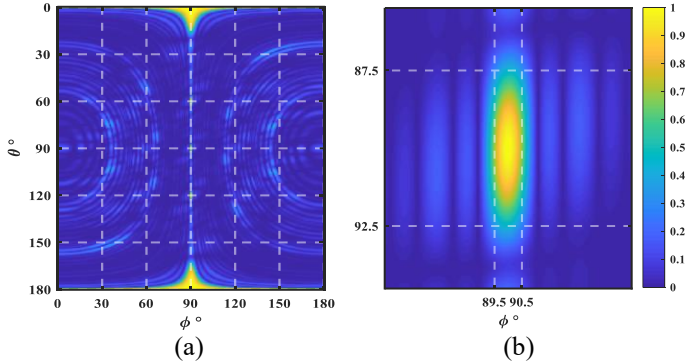

(a) (b)

Fig. 5. The resolution analysis of the designed array with (a) ambiguity function map and (b) detail of the ambiguity function map.

According to the simulation results of the ambiguity function, the resolution of the antenna array reaches up to 1 degree in the horizontal direction with ladder shape design and 5 degrees in the vertical direction. The resolution can be optimized to 3.8 degrees in vertical direction if the ladder shape design is used. However, the horizontal placement of the antennas is limited by the locations of the RF chips on the physical RF board which is shown in Figure 4(c). Then the horizontal antennas are arranged in uniform distribution. The improvement of two-dimensional resolution is the basis of point cloud accuracy and space detection ability, and the detailed design parameters are shown in TABLE I.

TABLE I
THE DESIGN PARAMETERS OF THE ANTENNA ARRAY

| Angle resolution | | FoV | |
|---|---|---|---|
| Azimuth direction | Elevation direction | Azimuth direction | Elevation direction |
| 1° | 5° | 180° | 30° |

## 3. Dynamic CFAR

The basic mechanism of target detection in RD maps is the judgement of target existence using CFAR method, which is shown in Figure 6. The decision thresholds $T_R$ and $T_D$ are calculated in range and Doppler dimensions with the judgement written as

$$J(x) = \begin{cases} \text{target}, & \text{if } x \geq T_R \text{ and } x \geq T_D \\ \text{noise}, & \text{otherwise} \end{cases} \quad (20)$$

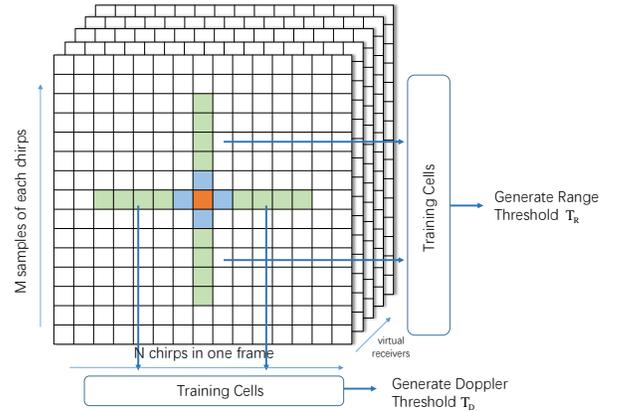

Fig. 6. Sampling points under detection in RD maps.

However, the target with weak reflection usually cannot be detected when the training cells appear high power, bringing the accuracy decrease of the spatial target detection. Therefore, the key point is to determine adequate edge and threshold for the detection.

In the case of the 2D target detection, CFAR with fixed parameters is always difficult to achieve more accurate performance when the environment is complicated with un-uniform distributed clutters. Thus, it is necessary to distinguish object in different environment clutters to perform sensitively detection and capture the detailed characteristics of the target accurately. The dynamic CFAR algorithm is proposed to tackle this problem. Firstly, the detection area is estimated from other sensors, such as camera or digital map, Then, the sliding window parameters of CFAR are determined using a prior knowledge of the clutter distribution, to determine which region the CUT belongs to. Consequently, the corresponding background powers is calculated according to the judgment result. Finally, the dynamic adjustment according to different thresholds strategies is taken to obtain accurate detections.

The modified dynamic CFAR can be used in the uniform distributed clutter environments or complicated environments with un-uniform distributed clutters. In the case of the complicated environment, the edge areas are identified refers to the reference window. Then the location region of the CUT is judged strong or weak clutter region. Finally, two detectors are introduced to perform target detection and the results of both area sides are merged. The detailed diagram is shown in Figure 7. The CUT is located in region 1, and the parameters of region 1 are used to calculate the detection threshold which are variance of the clutter power $\delta_1$ and the length $M'$, when variance of the clutter power $\delta_2$ and the length $N'$ present the parameters of region 2.

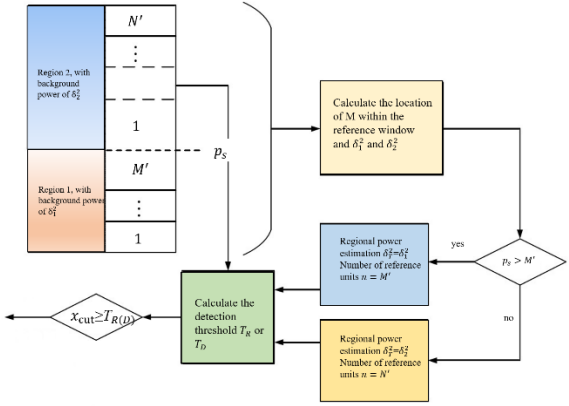

Fig. 7. The diagram of the proposed dynamic CFAR.

The maximum likelihood estimation method is used to determine the clutter edge position which can be described as:

$$L(x_1, x_2, \ldots, x_{M'+N'}|H_{K'}) = \prod_{m'=1}^{M'} \frac{1}{\delta_{1,K'}^2} \exp\left(\frac{x_{m'}}{\delta_{1,K'}^2}\right) \prod_{n'=1}^{N'} \frac{1}{\delta_{2,K'}^2} \exp\left(\frac{x_{n'}}{\delta_{2,K'}^2}\right) \quad (21)$$

where $x_1, x_2, \ldots, x_{M'+N'}$ are the sampling points of the RD map, $H_{K'}$ is the hypothesis of the clutter distribution.

With taking logarithms on both sides of (21)(21), the maximum likelihood estimation of the threshold can be written as:

$$\ln(L(x_1, x_2, \ldots, x_{M'+N'}|H_{K'})) = \sum_{m'=1}^{M'} \left(-\frac{x_{m'}}{\delta_{1,K'}^2} - \ln\delta_{1,K'}^2\right) + \sum_{n'=1}^{N'} \left(-\frac{x_{n'}}{\delta_{2,K'}^2} - \ln\delta_{2,K'}^2\right) \quad (22)$$

Abbreviate $L_{K'}$ for $L(x_1, x_2, \ldots, x_{M'+N'}|H_{K'})$ and find the gradient on both sides of (22), which is:

$$\begin{cases} \frac{\partial \ln L_{K'}}{\partial \delta_{1,K'}^2} = 0 \\ \frac{\partial \ln L_{K'}}{\partial \delta_{2,K'}^2} = 0 \end{cases} \quad (23)$$

The estimates of $\delta_{1,K'}^2$ and $\delta_{2,K'}^2$ can be obtained as

$$\begin{cases} \delta_{1,K'}^2 = \frac{1}{M'} \sum_{m'}^{M'} x_{m'} \\ \delta_{2,K'}^2 = \frac{1}{N'} \sum_{n'}^{N'} x_{n'} \end{cases} \quad (24)$$

Then the expression of $\ln L_{K'}$ can be arranged and rewritten as

$$\ln L_{K'} = -M' - M'\ln\delta_{1,K'}^2 - N' - N'\ln\delta_{1,K'}^2 \quad (25)$$

Ignoring the constants and the split position of area 1 and area 2 can be optimized and written as

$$p_s = \underset{K' \in (1,2,\ldots,M'+N')}{\arg\min} \left(M'\ln\delta_{1,K'}^2 + N'\ln\delta_{1,K'}^2\right) \quad (26)$$

With the determined split position, the SOCA-CFAR algorithm is introduced to detect the target existence. For the same target in different scenarios, the scattering field of radar wave can be simulated and analyzed by the same person as the target, which is shown in Figure 8. In different cases, the same person target, combined with different scenes, has different echo profiles with different noises. The same method produces different noise thresholds, which increase the difficulty of the detection accuracy. Based on the point cloud imagery process described above, the proposed method to estimate the boundaries of un-uniform regions using dynamic thresholds for different regions is considered to enhance the point cloud accuracy.

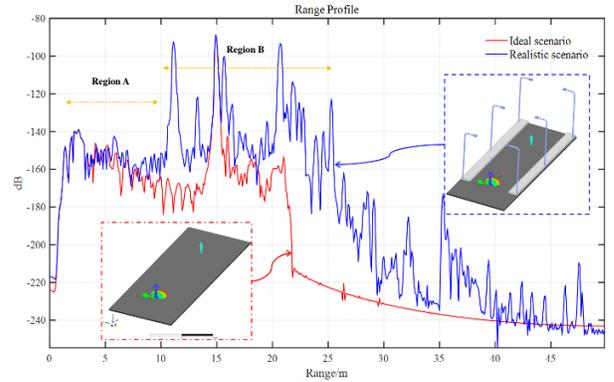

Fig. 8. Range profile comparison of ideal and realistic scenarios.

The corresponding threshold is selected according to the length of the reference cells, which is shown in Figure 9, where the target C can be detected using the dynamic CFAR proposed in this paper. When more point cloud information is acquired, the target can be detected more accurately.

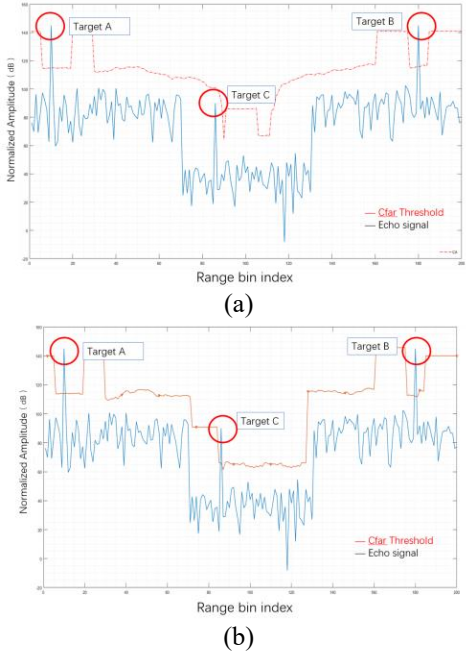

ig.9. Detection performance of (a) traditional CFAR and (b) dynamic CFAR proposed in this paper.

### C. Angular resolution optimization

As described in (12), the estimated angle $\alpha_q$ has the value $\alpha_q = \theta_q$ when the right part of the equation reaches the maximum value. However, the angle $\theta_q$ is unknown before extraction from the received signals.

$$\tilde{\alpha}_q = \arg\max_{\alpha_q} \left( \sum_{l=1}^{L} a(l) \cdot \exp\left(-jw_q(l-1)\right) \right) \quad (27)$$

where $a(l) = \exp(-j2\pi(l-1)d \sin\theta_q / \lambda)$ represents the echo signal of target from the $l$-th channel and $w_q = -2\pi d \sin\alpha_q / \lambda$ is the angular frequency sorted out.

The expression of the object function in (27)(27) has the similar formation of the discrete Fourier transform (DFT) with length $l$. When the spatial sampling space $d$ is fixed, the effective approach to improve the resolution of $\alpha_q$ is to increase the number of sampling length $l$, which means the deployment of more virtual antennas in the objective spatial dimension. In the antenna array design of this paper mentioned before, the synthesized array in horizontal direction with large number of virtual antennas is proposed and analyzed, with which the whole MIMO array is designed with small redundancy and high angular resolution in horizontal direction.

The shooting and bouncing rays (SBR)[17] tracing method is introduced to verify the angular resolution and spatial localization ability of the designed MIMO array. The simulation analysis of the designed array is shown in Figure 10. To deploy the simulation geometry as shown in Figure 10(a), the mmWave radar is located at one end of the road and has a certain height from the road plane. At the other end of the road, two reflective objects (such as metal pellets) are placed. In the former content of the antenna design, the antenna pattern has been computed using the EM simulator. By importing the radiation characteristics of each antenna according the actual array positions and solving the EM field distribution in the whole range of the mmWave radar detection using SBR tracing method, the EM characteristics of the detected targets can be obtained which is shown in Figure 10(b) and Figure 10(c).

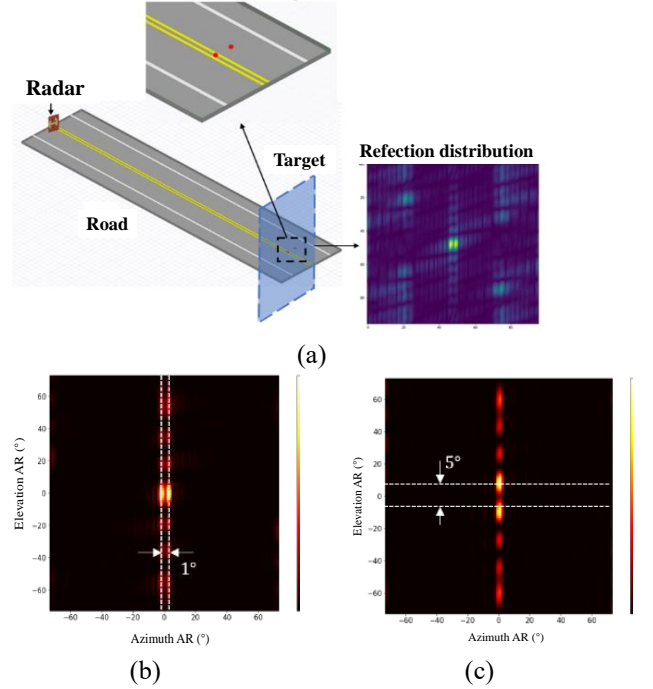

Fig.10. Simulation analysis of the designed MIMO array with (a) SBR tracing geometry and reflection distributions of two objects in (b) horizontal direction and (c) vertical direction.

The angular resolutions reach 1 degree and 5 degrees in horizontal direction and vertical direction respectively as the SBR tracing analysis shows, which is consistent with the ambiguity function analysis in former content of the antenna design and can both indicate the detection performance of the MIMO array. After the processing of the target echo signals, the reflection intensities with spatial distribution can be obtained. The numerical analysis shows that the designed MIMO array can distinguish the reflectors in high resolution which means high accuracy of the DOA estimation, making it possible to generate high-quality point cloud using the designed antenna array.

## IV. EXPERIMENT RRESULTS

### A. 3D spatial resolutions

Since the velocity accurate of the mmWave just depends on the parameters of pulse repetition frequency (PRF) and chirp number per frame, the 3D spatial resolutions in range, azimuth and elevation are analyzed. The range resolution analysis is performed and the result is shown in Figure 11. The designed range of the target is ~9 m and the measured profile shows a peak value in the distance of 8.99 m with high SNR.

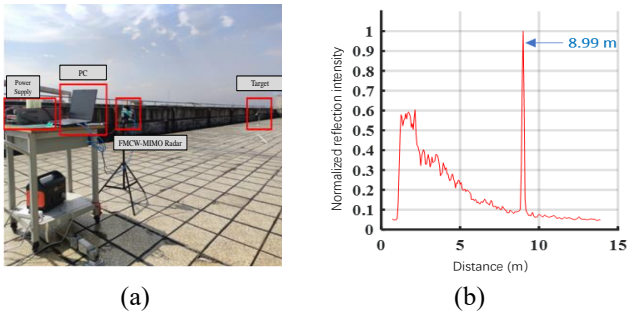

(a)　　　　　　　　　　(b)

Fig.11. Range resolution analysis with (a) the experiment scenario deployment and (b) the measured range profile of the target.

The angular resolutions in azimuth and elevation are tested in the anechoic chamber, as shown in Figure 12(a). Two metal objects located in the FoV of the mmWave radar act as the point target, which are placed in vertical or horizontal direction depending on the resolution test in azimuth or elevation. Figure 12(b) shows the measurement of the two targets with angular resolution 1.18 degrees in azimuth. Figure 12(c) shows the measurement of the two targets with angular resolution 5.9 degrees in azimuth.

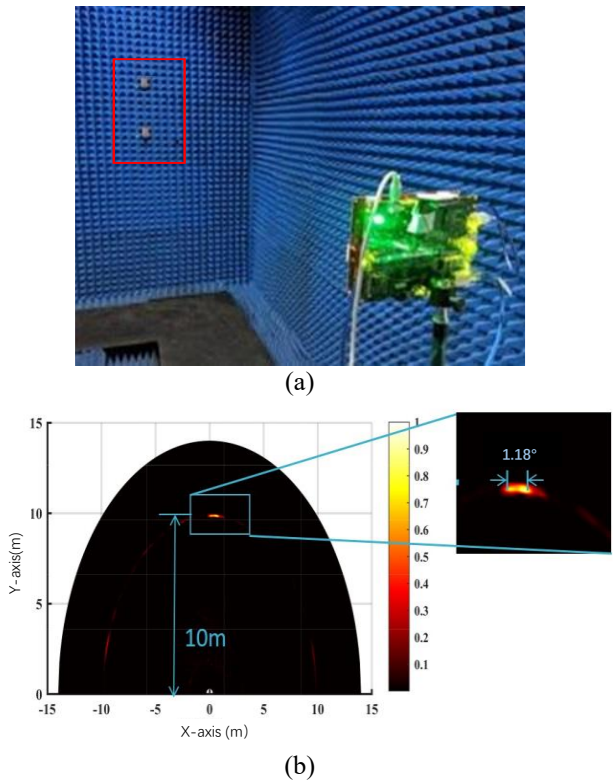

(a)

(b)

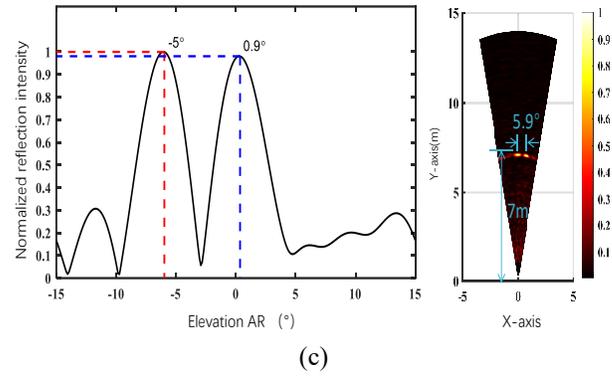

(c)

Fig.12. Angular resolutions test with (a) the experiment scenario deployment and the measurements of two targets (b) in azimuth and (c) in elevation.

### B. The detectability of weak target

Using the designed antenna, the detection performance of weak target is tested. The experiment scenario is setup on the rooftop with a wooden target, which is shown in Figure 13.

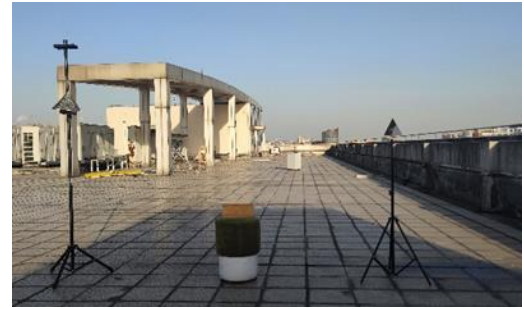

Fig.13. The experiment scenario deployment with weak target.

The on-chip antenna of comparison is the TI AWR2243 integrated in the mmWave cascade imaging radar RF evaluation module with the original manufacturer patch antenna. The optimized antenna designed in this paper has higher SNR and the weak target can be detected, which can be seen in Figure 14(b). Figure 14(a) shows the RD map generated using TI AWR2243, where the weak target cannot be perceived with the target signal level is very low.

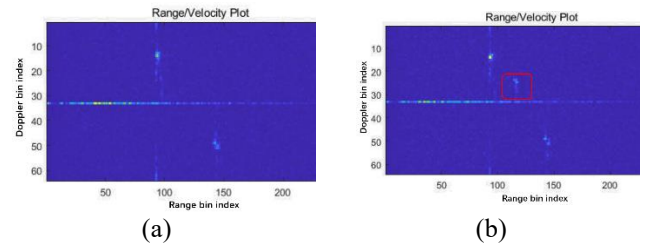

(a)　　　　　　　　　　(b)

Fig.14. Weak target detection using (a) the original manufacturer patch antenna of TI AWR2243 and (b) the optimized antenna designed in this paper.

### C. The performance of point cloud generation

The enhancement of the point cloud is verified in this section through 2 imagery cases of the roadside imagery and the stationary pedestrian imagery. The roadside imagery experiment results are shown in Figure 15. The comparison method is the original one of TI AWR2243 whose performance is shown in Figure 15(a). Figure 15(b) is the performance of the enhancement method proposed in this paper.

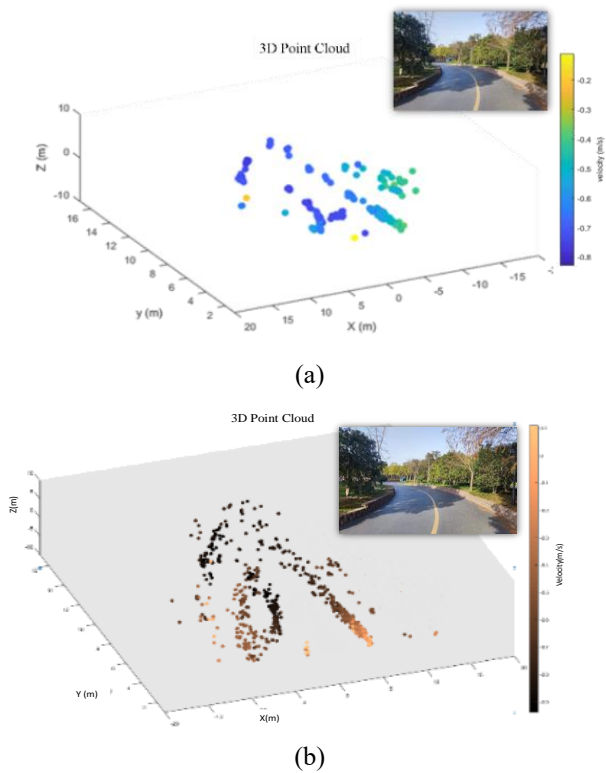

Fig.15. The roadside imagery using (a) the original method of TI AWR2243 and (b) the proposed method.

The detailed analysis is listed in TABLE II. The width accuracy is improved about 40% and the height accuracy is improved about 60% using the proposed method. The point cloud density is improved about 57%. The specific accuracy is only part of the great improvement. The distribution and the contour of the target are clearer, which supply high quality images to target recognition and environment perception.

TABLE II
THE COMPASION OF RQADSIDE IMAGERY PERFORMANCE

| Method | Width | Height | Number of points |
|---|---|---|---|
| Ground truth | 0.7m | 0.6m | / |
| Original | 0.49m | 0.35m | 210 |
| Proposed | 0.8m | 0.56m | 331 |

Another case is the stationary pedestrian imagery which is shown in Figure 16. The comparison method is still the original TI AWR2243 whose performance is shown in Figure 16(b), when the proposed one is shown in Figure 16(c).

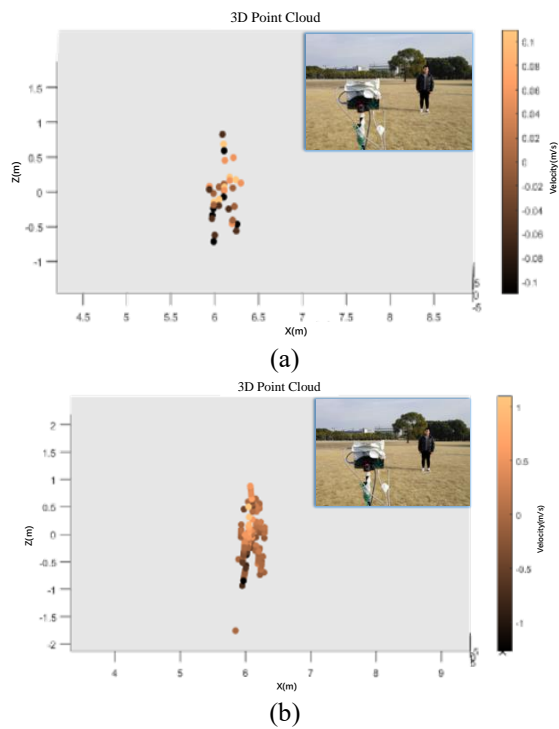

(a)

(b)

Fig. 16. The pedestrian imagery experiment (a) the original method of TI AWR2243 and (b) the proposed method.

It can be seen from the actual measurement, the point cloud generated using the proposed method has significant improvements in contour degree, with about 21%, 64% and 286% of point density increases in height, width and total density respectively.

TABLE III
THE COMPASION OF RQADSIDE IMAGERY PERFORMANCE

| Method | Length | Width | Height | Number of points |
|---|---|---|---|---|
| Ground truth | 0.7m | 0.15m | 1.82m | / |
| Original | 1.0m | 0.35m | 1.54m | 60 |
| Proposed | 0.8m | 0.23m | 1.8m | 232 |

## V. CONCLUSION

This paper proposes a point cloud enhancement method for high-precision 4D mmWave radar with the small redundancy MIMO array and dynamic CFAR. The ladder shape array is used in the design of MIMO array deployment to decrease the redundancy and improve the angular resolution. The dynamic CFAR enables dynamic adjustment of the judgment threshold by discriminating and calculating different scanning regions. Using the FT-based DOA algorithm, the designed radar system obtains higher regular resolution and more point cloud generation. To validate the method, we conducted experimental tests for different targets, such as road scenes and portraits. The results show that the enhanced point cloud imagery method has the improvement of 50% accuracy and 57% density using the same RF chips and detection parameters. These results demonstrate that the proposed method has better performance to be used in 4D mmWave radar imagery and is valuable for practical usages.

## REFERENCES

[1] Henry Alexander Ignatious, Hesham-El- Sayed, Manzoor Khan, An overview of sensors in Autonomous Vehicles, Procedia Computer Science, Volume 198, 2022, Pages 736-741, ISSN 1877-0509, https://doi.org/10.1016/j.procs.2021.12.315.(https://www.sciencedirect.com/science/article/pii/S1877050921025540)

[2] Y. Cheng, J. Su, M. Jiang and Y. Liu, "A Novel Radar Point Cloud Generation Method for Robot Environment Perception," in IEEE Transactions on Robotics, vol. 38, no. 6, pp. 3754-3773, Dec. 2022, doi: 10.1109/TRO.2022.3185831.

[3] M. Jiang et al., "4D High-Resolution Imagery of Point Clouds for Automotive mmWave Radar," in IEEE Transactions on Intelligent Transportation Systems, doi: 10.1109/TITS.2023.3258688.

[4] F. Engels, P. Heidenreich, M. Wintermantel, L. Stäcker, M. Al Kadi and A. M. Zoubir, "Automotive Radar Signal Processing: Research Directions and Practical Challenges," in IEEE Journal of Selected Topics in Signal Processing, vol. 15, no. 4, pp. 865-878, June 2021, doi: 10.1109/JSTSP.2021.3063666.

[5] Z. Xia and F. Xu, "Time-Space Dimension Reduction of Millimeter-Wave Radar Point-Clouds for Smart-Home Hand-Gesture Recognition," in IEEE Sensors Journal, vol. 22, no. 5, pp. 4425-4437, 1 March1, 2022, doi: 10.1109/JSEN.2022.3145844.

[6] B. Tan et al., "3-D Object Detection for Multiframe 4-D Automotive Millimeter-Wave Radar Point Cloud," in IEEE Sensors Journal, vol. 23, no. 11, pp. 11125-11138, 1 June1, 2023, doi: 10.1109/JSEN.2022.3219643.

[7] Human–vehicle classification using feature-based SVM in 77-GHz automotive FMCW radar. Seongwook Lee, Young-Jun Yoon, Jae-Eun Lee, Seong-Cheol Kim. First published: 11 August 2017. https://doi.org/10.1049/iet-rsn.2017.0126

[8] Hyun, E.; Jin, Y.-S.; Lee, J.-H. A Pedestrian Detection Scheme Using a Coherent Phase Difference Method Based on 2D Range-Doppler FMCW Radar. Sensors 2016, 16, 124. https://doi.org/10.3390/s16010124

[9] E. Hyun, "Moving and stationary target detection scheme using coherent integration and subtraction for automotive FMCW radar systems," IEEE Radar Conf., pp. 0476-0481. Jun. 2017,

[10] E. Hyun, "A pedestrian detection scheme using a coherent phase difference method based on 2D range-doppler FMCW radar," MDPI Sensors, vol. 16, no. 124, Jan. 2016.

[11] J. Bai, "Traffic participants classification based on 3D radio detection and ranging point clouds," IET Radar Sonar Navigation, vol. 16, no. 2, pp. 278-290, Oct. 2021.

[12] S. M. Patole, M. Torlak, D. Wang and M. Ali, "Automotive radars: A review of signal processing techniques," in IEEE Signal Processing Magazine, vol. 34, no. 2, pp. 22-35, March 2017, doi: 10.1109/MSP.2016.2628914.

[13] Mark Richards, Fundamentals of Radar Signal Processing, McGraw Hill, 2005

[14] Li X, Wang W, Wang X ,et al.Synthetic sparse planar array design for two-dimensional DOA estimation[J].Digital Signal Processing, 2022(120-).

[15] A. Ahmed and Y. D. Zhang, "Generalized Non-Redundant Sparse Array Designs," in IEEE Transactions on Signal Processing, vol. 69, pp. 4580-4594, 2021, doi: 10.1109/TSP.2021.3100977.

[16] Q. JinYou, Z. JianYun and L. ChunQuan, "The Ambiguity Function of MIMO Radar," 2007 International Symposium on Microwave, Antenna, Propagation and EMC Technologies for Wireless Communications, Hangzhou, China, 2007, pp. 265-268, doi: 10.1109/ MAPE. 2007.4393596.

[17] Stephen Kasdorf, Blake Troksa, Branislav M. Notaroš, "Some Advances in Shooting-Bouncing-Rays Asymptotic Propagation Methodologies", 2021 International Applied Computational Electromagnetics Society Symposium (ACES), pp.1-3, 2021.

[18] https://www.ti.com/tool/MMWCAS-RF-EVM